\documentclass[aps,prl,showpacs,floatfix,superscriptaddress,twocolumn,color]{revtex4}
\usepackage{graphicx}
\usepackage{amsmath}
\usepackage{amsfonts}
\usepackage{amssymb}
\usepackage{wasysym}
\usepackage{dsfont}
\usepackage{color}
\usepackage[normalem]{ulem}

\definecolor{redmarker}{rgb}{0.9,0.0,0.0}
\definecolor{greenmarker}{rgb}{0.0,0.9,0.0}
\definecolor{bluemarker}{rgb}{0.1,0.1,0.9}

\begin{document}

\title{Collective modes of coupled phase oscillators with delayed coupling}

\author{Sa\'ul Ares}			
\altaffiliation[Present address: ] {Logic of Genomic Systems Laboratory, Centro Nacional de Biotecnolog\'ia - CSIC. Calle Darwin 3, 28049 Madrid, Spain}
\affiliation{Max Planck Institute for the Physics of Complex Systems, N\"othnitzer Str. 38, 01187 Dresden, Germany}
\affiliation{Grupo Interdisciplinar de Sistemas Complejos (GISC), Spain}
\author{Luis G. Morelli}		
\affiliation{Max Planck Institute for the Physics of Complex Systems, N\"othnitzer Str. 38, 01187 Dresden, Germany}
\affiliation{Max Planck Institute of Molecular Cell Biology and Genetics, Pfotenhauerstr. 108, 01307 Dresden, Germany}
\affiliation{CONICET, Departamento de F\'isica, UBA, Ciudad Universitaria,  1428 Buenos Aires, Argentina}
\author{David J. J\"org}		
\affiliation{Max Planck Institute for the Physics of Complex Systems, N\"othnitzer Str. 38, 01187 Dresden, Germany}
\author{Andrew C. Oates}		
\affiliation{Max Planck Institute of Molecular Cell Biology and Genetics, Pfotenhauerstr. 108, 01307 Dresden, Germany}
\author{Frank J\"ulicher}		 \email{julicher@pks.mpg.de}
\affiliation{Max Planck Institute for the Physics of Complex Systems, N\"othnitzer Str. 38, 01187 Dresden, Germany}
\date{\today}

\begin{abstract}  
\noindent 
We study the effects of delayed coupling on timing and pattern formation in spatially extended systems of dynamic oscillators.
Starting from a discrete lattice of coupled oscillators, we derive a generic continuum theory for collective modes of long wavelength. 
We use this approach to study spatial phase profiles of cellular oscillators in the segmentation clock, a dynamic patterning system of vertebrate embryos.
Collective wave patterns result from the interplay of coupling delays and moving boundary conditions.
We show that the phase profiles of collective modes depend on coupling delays.
\end{abstract}
\maketitle

\noindent 
In complex dynamical systems interactions
between different elements can give rise to dynamical order and spatio-temporal patterning \cite{winfree,manrubia,pikovsky}. 
These interactions can themselves be the result of a complex process. Therefore
coupling can have internal dynamics that involve time delays.
The importance of time delays in the coupling of oscillators 
was first recognized by Schuster and Wagner~\cite{schuster89},
who showed that two oscillators can entrain even if coupling is delayed. In this case, 
 multistability of dynamic states can occur as a consequence
of time delays and the collective frequency of the system depends on the delay time. 
The roles of time delays have been addressed in studies of different oscillator systems. 
It was found that
coupling delays can give rise to a rich variety of behaviors \cite{niebur91,yeung99,zanette00,jeong02,sethia08,sethia10,eguiluz11,montbrio06}. 

Significant time delays in the coupling of oscillators occur in many systems in
biology, engineering and physics. Coupling delays between a sending and a receiving
element can arise due to 
(i) intrinsic times of signal generation in the sending element,
(ii) the finite propagation velocity of signals, and
(iii) the slow signal processing of the receiving element.
Coupling delays of type (ii) are inevitable in some engineered systems
\cite{wunsche05,pogorzelski08a,pogorzelski08b,wu96,hale94a} and neuronal systems \cite{nishii,ermentrout09,kim97,li10}.
In some cases such delays can be used to implement control schemes~\cite{choe10,fiedler10}.
In the context of signaling processes between biological cells \cite{lewis03,veflingstad05,momiji09},
coupling delays of type (i) and (iii) occur naturally because 
of the complex internal kinetics of intercellular signaling.

An important biological example in which the delayed coupling of dynamic
oscillators plays a key role for the formation of patterns is  the the so-called segmentation clock \cite{roellig11,pourquie11}.
This system operates during embryonic development of all vertebrate animals,
generating a segmented morphology along the vertebrate body axis of the embryo. 
These segments, called somites, are the embryonic precursors of adult vertebrae.
Somites are formed sequentially in a dynamic tissue which elongates during the process.
The segmentation clock is thus a rhythmic pattern generator in the tissue resulting from the collective organization of many cells. 
Each cell is an autonomous oscillator with time-periodic activation of certain genes~\cite{masamizu06,lewis03}.
These oscillators are noisy and are coordinated by intercellular signaling~\cite{riedel07}.

It was recently suggested that coupling delays between genetic oscillators play an important role
for the dynamics of the segmentation process and that they influence the collective frequency
of cellular oscillations \cite{morelli09}. By comparing theory with quantitative experiments in fish embryos
it was subsequently shown that coupling delays indeed play a crucial role in pattern formation 
and that the collective oscillation frequency is altered in fish with mutations affecting the coupling process
between oscillators \cite{herrgen10}.  The segmentation clock is therefore a prime example of a population of coupled
oscillators in which coupling delays play a crucial role. Its understanding requires a theory
of locally coupled oscillators with delays in a spatially extended system.

Despite their general interest, the effects of coupling delays in spatially extended systems 
are still poorly understood. It has been shown, using a low dimensional approximation \cite{ott08},
that such systems can display a wide range of spatiotemporal patterns \cite{lee11}.
An important approach to describe spatially extended systems is to focus on long-wavelength
modes, which can be described by a simplified 
continuum limit which ignores details on small scales.
 Continuum descriptions have been developed for chains of coupled phase oscillators without delay
\cite{kuramoto84,kopell86} and for rings of coupled phase oscillators in the limit of short delay times
\cite{pogorzelski08b}. It was shown that  short time delays are equivalent to a phase shift in the coupling 
\cite{ermentrout94,izhikevich98}. This simplification is lost when coupling 
delays are longer than the time scale defined by the coupling strength. In this case
time delays  have to be considered explicitly \cite{izhikevich98}.

In this Letter, we study the collective oscillatory modes of the vertebrate segmentation clock.
We introduce a generic continuum description for the long wavelength modes 
of spatially extended lattices of coupled oscillators with arbitrary coupling delays.
We use this theory to discuss the emergent collective frequency, as well as phase profiles of collective modes. 

The vertebrate segmentation clock is a tissue level pattern generator~\cite{roellig11}.
It consists of a population of cells that collectively generate oscillating activity patterns of cyclic genes in the tissue. 
Neighboring cells are coupled via a slow molecular signaling system that introduces time delays \cite{morelli09,herrgen10}.
Thus, we consider a system of coupled oscillators defined on a regular $d$-dimensional hypercubic lattice with lattice spacing $a$.
Oscillators sit on lattice sites with position ${\bf x}_i=a{\bf i}$
with ${\bf i}=(i_1,..,i_d)$, $i_n\in \mathbb{Z}$.
The evolution equation for the phase $\theta_{\bf i}$ of the oscillator ${\bf i}$, coupled to its nearest neighbors ${\bf j}$, is
\begin{equation}\label{eq.Ddim}
\frac{d\theta_{\bf i}(t)}{dt}=\omega_{\bf i} + \frac{\varepsilon_{\bf i}}{2d}\sum\limits_{|{\bf j}-{\bf i}|=1} h(\theta_{\bf j}(t-\tau)-\theta_{\bf i}(t)) \ ,
 \end{equation}
where $\omega_{\bf i}$ is the intrinsic frequency of the oscillator ${\bf i}$,
$\varepsilon_{\bf i}$ denotes the coupling strength of this oscillator to its neighbors,
and $\tau>0$ is a time delay in the coupling.
The coupling is described by the $2\pi$-periodic function $h$.

We are interested in long wavelength collective modes, for which the phases $\theta_{\bf i}$ 
vary smoothly over distances that are long compared to the lattice spacing $a$~\cite{kuramoto84}.
We assume that intrinsic frequency and coupling vary smoothly in space so as to fulfill this condition.
In this situation, we can approximate the phases of the oscillators in the discrete system by a continuum phase field $\theta({\bf x},t)$.
In the following we derive an equation for $\theta({\bf x},t)$, which describes the long wavelength dynamics of the discrete oscillator system, Eq.~(\ref{eq.Ddim}).
To simplify the notation, we derive the continuum theory on a one-dimensional lattice.
The generalization to higher dimensions is straightforward.
We first perform a Taylor expansion of the coupling function $h$ in Eq. (\ref{eq.Ddim}) in powers of $a$, assuming
that $a/\ell\ll 1$, where $\ell\simeq (\partial \theta/\partial x)^{-1}$ is a characteristic wavelength of the collective mode,
\begin{align} \label{eq.taylorx}
h(\Delta_\tau^\pm) &= h(\Delta_\tau)
\pm h'(\Delta_\tau) \frac{\partial \theta_\tau}{\partial x} a \\
&+\frac{1}{2} \bigg[h''(\Delta_\tau) \left(\frac{\partial \theta_\tau}{\partial x} \right)^2 
+h'(\Delta_\tau) \frac{\partial^2 \theta_\tau}{\partial x^2} \bigg]a^2 +  \mathcal{O}(a^3)  \, . \nonumber
\end{align}
The prime denotes the derivative of $h$ with respect to its argument,
and we have defined
\begin{align}
\Delta_\tau^\pm(x,t) &\equiv \theta(x \pm a,t-\tau)-\theta(x,t), \\
\Delta_\tau(x,t) &\equiv \theta(x,t-\tau)-\theta(x,t),	\\
\theta_\tau(x,t) &\equiv \theta(x,t-\tau)			\, .
\end{align}
Introducing continuum functions $\omega(x)$ and $\varepsilon(x)$ of the intrinsic frequency and coupling strength of the oscillators,
together with Eq.~(\ref{eq.taylorx}), we obtain a continuum description. The  phase field obeys
\begin{align}
\begin{split}
\frac{\partial\theta(x,t)}{\partial t} &=\omega(x) + \varepsilon (x) h(\Delta_\tau(x,t))\\
&\quad + \frac{\varepsilon (x) a^2}{2} h''(\Delta_\tau(x,t))\left(\frac{\partial \theta(x,t-\tau)}{\partial x}\right)^2  \\
&\quad + \frac{\varepsilon (x) a^2}{2} h'(\Delta_\tau(x,t))\frac{\partial^2 \theta(x,t-\tau)}{\partial x^2}  \ . 
\end{split}\label{eq.ca1d}
\end{align}
Note that the first order contribution in $a$ vanishes.
Higher order terms can be neglected in the long wavelength limit and have been dropped here.

For arbitrary dimension we find
\begin{align}
\begin{split}
&\frac{\partial\theta ({\bf x},t)}{\partial t} =\omega({\bf x}) +
\varepsilon ({\bf x}) h(\theta({\bf x},t-\tau)-\theta({\bf x},t))\\
&\quad + \frac{\varepsilon ({\bf x}) a^2}{2d} \,\, h'' \left( \theta({\bf x},t-\tau)-\theta({\bf x},t) \right) \,\, \left[\nabla \theta({\bf x},t-\tau)\right]^2  \\
&\quad + \frac{\varepsilon ({\bf x}) a^2}{2d} \,\, h' \left( \theta({\bf x},t-\tau)-\theta({\bf x},t) \right) \,\, \nabla^2 \theta ({\bf x},t-\tau)  \ ,
\end{split}\label{eq.caDd}
\end{align}
where ${\bf x} \in \mathds{R}^d$ is a position vector in $d$-dimensional space.
The coupling delay $\tau$ enters in the spatial derivatives of the phase field, as well as 
in the arguments of the coupling function and its derivatives.
For a vanishing delay, $\tau=0$, we recover the classical case of locally coupled oscillators~\cite{kuramoto84}.
Eq.~(\ref{eq.caDd}) describes the collective modes of general extended systems of oscillators with coupling delays.

\begin{figure}[t]
\begin{center}
\includegraphics[width=6.5cm]{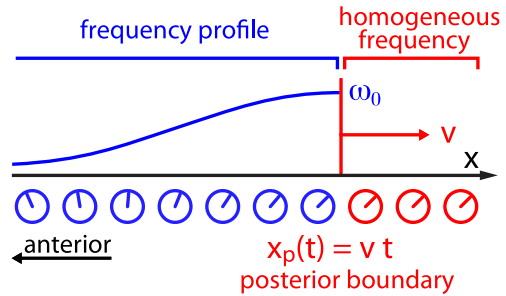}
\caption{One dimensional lattice of coupled oscillators representing the segmentation clock with a frequency profile (blue) and posterior boundary at position $x_p$ moving with velocity $v$ (red). The oscillators self-organize in a phase pattern.
The frequency profile starts at the posterior boundary with value $\omega_0$ and drops to zero as $x\to-\infty$.}
\label{fig.scheme}
\end{center}
\end{figure}

We now employ our approach to describe gene activity patterns of cells 
in the segmentation clock of vertebrate embryos. 
We simplify our description of the segmentation process by using 
a semi-infinite one-dimensional geometry, Fig.~\ref{fig.scheme}, in which
a system of oscillators is described by Eq.~(\ref{eq.ca1d}). 
The system in which oscillators create patterns
extends from the anterior for $x\rightarrow -\infty$
to a posterior boundary at $x_p(t)= x_0+vt$. This moving boundary describes the
elongating tip of the tissue. Here  $v$ is  the elongation velocity of the 
tissue and $v/a$ is the rate at which cells are added at the extending end.
In the following we choose $x_0=0$ without
loss of generality.

In the segmentation clock, concentration gradients of signaling molecules exist across the tissue.
The source of these gradients is located at the tip of the elongating tissue, and the resulting concentration gradients move together with the tip.
Perturbations of these molecular gradients produce effects consistent with alteration of the intrinsic frequency of the cellular oscillators \cite{dubrulle01,sawada01}.
Motivated by these observations, the spatial profiles $\omega(x)$ and $\varepsilon(x)$ are assumed to depend only 
on the distance $d=x_p-x$ from the elongating tip~\cite{palmeirim97,kaern00,jaeger01,giudicelli07,morelli09},
i.e.~they travel together with the moving boundary.
Therefore, we introduce a reference frame co-moving with the tip boundary, where the frequency and coupling strength profiles are stationary.
We define the coordinate $y=x-vt$ and the co-moving phase field $\vartheta(y,t) = \theta(y+vt,t)$.
Since $d=x_p-x=-y$, the functions $\varepsilon(y)$ and $\omega(y)$ are time-independent. 
We assume that these functions take maximal
values $\omega(0)=\omega_0$ and $\varepsilon(0)=\varepsilon_0$ at $y=0$, 
that they decay monotonically and vanish in the limit $y \rightarrow -\infty$. 

The fact that both $\varepsilon(y)$ and $\omega(y)$
vanish for large negative $y$ implies, according to Eq.~(\ref{eq.ca1d}), that the pattern $\theta(x,t)$ becomes stationary
for $x \rightarrow -\infty$, i.e.~$\partial \theta/\partial t\simeq 0$ in this limit. 
This stationary pattern describes the developing segments.
Pattern forming solutions can be obtained by the ansatz:
\begin{equation}\label{eq.stedans}
\vartheta(y,t)=\phi(y)+\Omega t  \, .
\end{equation}
Using this expression, the differential equation for the phase profile reads for $y<0$
\begin{align}
\begin{split}
\Omega	& = v \phi'(y) + \omega(y) + \varepsilon(y) h( \phi(y+v\tau) - \phi(y) -\Omega\tau) \Big. \\	
		& + \frac{\varepsilon(y) a^2}{2} h''\left( \phi(y+v\tau) - \phi(y) -\Omega\tau \right) (\phi'(y+v\tau))^2 \\
		& + \frac{\varepsilon(y) a^2}{2} h'\left( \phi(y+v\tau) - \phi(y) -\Omega\tau \right) \phi''(y+v\tau) \ .				
\end{split}\label{eq.ca1d_bis3}
\end{align}
The first term in Eq. (\ref{eq.ca1d_bis3}) 
is a drift describing phase transport due to the motion of oscillators in the co-moving system.
The moving boundary and the co-moving profiles of $\varepsilon$ and $\omega$,
together with coupling delays, introduce non-local effects in Eq.~(\ref{eq.ca1d_bis3}). 

Eq. (\ref{eq.ca1d_bis3}) is solved imposing boundary conditions at the moving tip, $y=0$. Because of the
nonlocal terms involving oscillators at $y+v\tau$ in Eq.~(\ref{eq.ca1d_bis3}), the boundary values
of $\phi(0)$, $\phi'(0)$ and $\phi''(0)$ are not sufficient to determine the solution. 
Rather, it is necessary to specify the function $\phi(y)$ in the interval $0\leq y\leq v\tau$. 
This nonlocal boundary condition reflects the fact that the history 
of oscillators that are attached to the end is important. 
This history is specified by this boundary condition.
We choose $\phi(y)=\phi_0$ for $y\geq 0$ as well as $\phi'(0)=0$ and $\phi''(0)=0$. 
This choice corresponds to the
assumption that all oscillators that enter the system at the tip oscillate with intrinsic frequency
$\omega_0$ and are coupled with strength $\varepsilon_0$, 
being in phase with their neighbors, Fig.~\ref{fig.scheme} (red).

Eq.~(\ref{eq.ca1d_bis3}) together with these boundary conditions imply that the collective frequency $\Omega$ obeys
\begin{equation}\label{eq.ca1d_bis8}
\Omega = \omega_0 + \varepsilon_0 h\left(  -\Omega\tau \right).				
\end{equation}
This transcendental equation is known to determine the collective frequency $\Omega$ in a population of identical oscillators. 
It can have one or several coexisting solutions for the collective frequency~\cite{schuster89,yeung99,earl03}.

The solutions of Eq.~(\ref{eq.ca1d_bis3}) with the boundary conditions imposed here,
describe a dynamic biological pattern generator that oscillates with collective frequency $\partial\vartheta/\partial t = \Omega$
and produces a stationary spatially periodic structure of length $S$ described by $\sin(\theta(x))\simeq \sin(2\pi x/S )$.
It is formed by the collective mode of coupled oscillators that corresponds to traveling waves 
induced at the moving tip that propagate towards the anterior where they slow down and stop.
From Eq. (\ref{eq.ca1d_bis3}), we find $\phi(y) \simeq y\Omega/v$ in the limit $y\rightarrow -\infty$
in which both $\omega$ and $\varepsilon$ vanish, implying 
\begin{equation}\label{eq.S}
S = vT \, ,
\end{equation}
where $T=2\pi/\Omega$ is the collective period.
This result shows that we recover the basic property of a general clock and 
wavefront mechanism of vertebrate segmentation~\cite{cooke76,roellig11}. 
The segment length $S$ depends only on the elongation velocity 
and the collective frequency of the oscillators at the posterior boundary.
It is independent of the shapes of the profiles of frequency and coupling strength.

\begin{figure}[t]
\begin{center}
\includegraphics[width=3.1in]{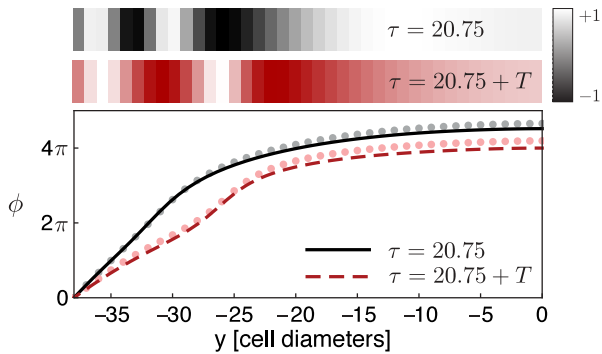}
\caption{Phase profiles of coupled oscillators with delayed coupling for the segmentation oscillator.
Upper panel: plots of $\sin\phi(y)$ corresponding to the phase profiles in the lower panel,
illustrating cyclic gene expression patterns in vertebrate segmentation.
Both simulations have the same collective period $T$ and elongation velocity $v$,
making the resulting segment length $S$ equal in both.
Lower panel: steady state phase profiles $\phi(y)$ in the co-moving frame 
obtained from the continuum approximation Eq.~(\ref{eq.ca1d_bis3}) (lines),
and numerical simulations of the discrete model given by Eq.~(\ref{eq.Ddim}) (dots). 
We use parameters previously determined for the zebrafish segmentation clock \cite{herrgen10}, 
resulting in a collective period $T=23.5$~min for both cases.
The coupling function is $h(\Delta)=\sin(\Delta)$.
Boundary condition is $\phi(y)|_{y\geq 0}=\mathrm{const.}$ 
The time units are minutes and the unit of length $a$ corresponds to one cell diameter, $a\approx10~\mu$m.
Only every second discrete oscillator is displayed, and the arrested oscillators to the left of $y=-39$ are not shown.
Parameters are:
$v=0.249$~cell~diameters/min, $\varepsilon(y)=\varepsilon_0$ for $y \geq -39$, with $\varepsilon_0=0.07$ min$^{-1}$, 
$\varepsilon(y)=0$ min$^{-1}$ for $y < -39$, 
$\omega(y)=\omega_0(1-e^{-(y+39)/27})/(1-e^{-39/27})$ for $-39 \leq y \leq 0$, with $\omega_0=0.2205$ min$^{-1}$,
$\omega(y)=0$ min$^{-1}$ for $y < -39$ and $\omega(y)=\omega_0$ for $y > 0$.
The biological motivation of the choice of $\varepsilon(y)$ and $\omega(y)$ is discussed in %
\cite{morelli09}.}
\label{fig.phaseprofiles}
\end{center}
\end{figure}

Solutions to Eq.~(\ref{eq.ca1d_bis3}) for boundary conditions given above are displayed
in Fig.~\ref{fig.phaseprofiles} (lines) for two different time delays.
As an example, realistic parameters previously determined for the zebrafish segmentation clock have been used, including typical profiles of the functions $\varepsilon(x)$ and $\omega(x)$~\cite{herrgen10}.
These profiles and parameter values are given in the caption of Fig.~\ref{fig.phaseprofiles}.
These solutions are compared to
simulations of the discrete oscillator system, Eq.~(\ref{eq.Ddim}), with the same parameters (dots). 
The comparison shows that our continuum theory provides an excellent approximation
for the behavior of the discrete system.
We have checked that
small perturbations to the discrete patterns 
relax back to the time-periodic states shown in Fig.~\ref{fig.phaseprofiles},
indicating that these patterns are locally stable.

Fig.~\ref{fig.phaseprofiles} also highlights the role of coupling delays in shaping the phase patterns.
The solid black line shows the phase profile for a time delay of $\tau=20.75$ min \cite{herrgen10}, 
while the dashed line was obtained for $\tau=44.25$ min keeping other parameters the same.
Thus, the value of $\tau$ in both cases differs by the collective period $T=23.5$ min.
Note that Eq.~(\ref{eq.ca1d_bis8}), which determines~$T$,  
is invariant under the transformation $\tau\to\tau+m T$ 
for any integer $m$. As a consequence, the two systems shown in Fig. \ref{fig.phaseprofiles} 
oscillate with the same period.
However, the phase profiles of these solutions differ 
because Eq.~(\ref{eq.ca1d_bis3}) is not invariant under this transformation. 
This effect may offer a way
to distinguish between different coupling delays
producing the same period in the segmentation clock, 
by means of phase profile data.
The shape of the observed pattern does not change when we introduce a white noise term to the discrete model, Eq.~(\ref{eq.Ddim}).
%
%

Our theory can account for several observed properties of the segmentation clock.
First, we showed that the simple clock and wavefront relation $S=vT$~\cite{cooke76,roellig11,schroeter10}
between segment length $S$, elongation velocity $v$ and oscillation period $T$,
holds under very general conditions of frequency profiles, time delays and coupling functions.
Second, our theory produces phase patterns that can quantitatively fit gene expression patterns observed in the segmentation clock \cite{morelli09,herrgen10}, see Fig.~\ref{fig.phaseprofiles}.
Third, independently of the details of the frequency and coupling strength profiles and the shape of the coupling function, our theory predicts through Eq.~(\ref{eq.ca1d_bis8}) the dependence of the collective frequency on coupling strength and coupling delay, consistent with experiments~\cite{herrgen10}.

Delays in oscillator coupling play an important role for the stability of synchronized phase profiles, as has been shown for simple cases \cite{earl03}.
Stable solutions of Eq.~(\ref{eq.caDd}) describe long wavelength modes of the coupled oscillator system given by Eq.~(\ref{eq.Ddim}).
The stability of these modes depends on time delays and boundary conditions.
Instabilities of Eq.~(\ref{eq.caDd}) respect to short wavelengths describe situations where the continuum description breaks down.

Systems of coupled oscillators can display individual variability~\cite{kuramoto,montbrio11} and be subject to dynamic fluctuations~\cite{sakaguchi88}.
Individual variability can be described by quenched disorder in the parameters, such as the frequency or coupling~\cite{kuramoto}.
Dynamic fluctuations can be described by an additional noise term in Eq.~(\ref{eq.caDd})~\cite{sanmiguel00}.
With the addition of this noise term, Eq.~(\ref{eq.caDd}) is a generalization of 
the Kardar--Parisi--Zhang  equation~\cite{kardar86} to time-delayed and inhomogeneous systems.
In the Kardar--Parisi--Zhang equation the interplay of noise and nonlinearity induces a rich phenomenology.
This indicates that the addition of noise to our problem
may have interesting effects on the dynamics.
Finally, fluctuations in the coupling delays could be straightforwardly accounted for in the theory, extending Eq.~(\ref{eq.caDd}) to include 
distributed coupling delays~\cite{macdonald}.

In this Letter, we have introduced a continuum description of long wavelength modes 
in extended systems of oscillators with coupling delays, Eq.~(\ref{eq.caDd}).
We have applied this continuum description to a problem from biology, the segmentation clock. 
This problem involves moving boundaries, which together with time delays give rise to nonlocal effects, see Eq. (\ref{eq.ca1d_bis3}).
We have proposed here that time delays have a key role in shaping the pattern of spatial phase profiles,
apart from setting the period of the segmentation clock.
The biological example of the segmentation clock shows that the continuum description is a 
powerful method to study extended systems with coupling delays.

%
We thank Bernold Fiedler and his group for warm epic discussions, Gautam Sethia for discussion on nonlocal coupling~\cite{sethia08,sethia10}, Lucas Wetzel and the Oates Lab.
S. A. acknowledges funding from Ministerio de Ciencia e Innovaci\'on (Spain) through grant MOSAICO. L. G. M. acknowledges ANPCyT PICT 876. L. G. M. and A. C. O. were supported by the Max Planck Society and the European Research Council under the European Communities Seventh Framework Programme (FP7/ 2007-2013)/ERC Grant No. 207634. S. A. and L. G. M. contributed equally to this work.


\end{document}